\documentclass[prd,aps,twocolumn,nofootinbib,preprintnumbers,superscriptaddress,preprintnumbers,balancelastpage,longbibliography]{revtex4-2}
\usepackage[english]{babel}

\usepackage{amsmath,mathtools,physics,xfrac}
\usepackage{graphicx}
\usepackage{afterpage}
\usepackage{float}
\usepackage{subfigure}
\usepackage{rotating}
\usepackage{multirow}
\usepackage{tabularx}
\usepackage{booktabs}
\usepackage{fancyhdr}
\usepackage[utf8]{inputenc}
\usepackage{theorem}
\usepackage{moreverb}
\usepackage{euscript}
\usepackage{psfrag}
\usepackage{slashed}
\usepackage{mathtools}
\usepackage{makecell}
\usepackage{adjustbox}
\usepackage{dcolumn}
\usepackage{bm}
\usepackage[dvipsnames]{xcolor}
\usepackage{graphics}
\usepackage{hyperref}
\usepackage{chngcntr}
\usepackage{aas_macros}
\usepackage{color,xcolor}
\hypersetup{
     colorlinks   = true,
     citecolor    = Aquamarine,
     urlcolor     = Aquamarine,
     linkcolor    = Aquamarine
}

\definecolor{darkblue}{rgb}{0.0,0.0,0.75}
\definecolor{darkred}{rgb}{0.6,0.0,0}
\definecolor{darkgreen}{rgb}{0.0,0.6,0.}
\newcommand{\ad}[1]{{\color{blue} #1}}

\counterwithout{equation}{section}

\newcommand{\eV}{~\mathrm{eV}}
\newcommand{\keV}{~\mathrm{keV}}

\newcommand{\species}{MVDM}
\newcommand{\lcdm}{$\Lambda$CDM}
\newcommand{\dz}{\Delta z}
\newcommand{\neff}{N_\mathrm{eff}}
\newcommand{\dneff}{\Delta N_\mathrm{eff}}

\newcommand{\rhosp}{\rho_\mathrm{\species}}
\newcommand{\Pd}{P_\mathrm{1D}}

\begin{document}
\title{\boldmath Mass varying dark matter and its cosmological signature}

\author{Anirban Das}
\email{anirbandas@snu.ac.kr}
\affiliation{Center for Theoretical Physics, Department of Physics and Astronomy, Seoul National University, Seoul 08826, South Korea}

\author{Subinoy Das}
\email{subinoy@iiap.res.in}
\affiliation{Indian Institute of Astrophysics, Bengaluru, Karnataka 560034, India}

\author{Shiv K. Sethi}
\email{sethi@rri.res.in}
\affiliation{Raman Research Institute, C.~V. Raman Avenue, Sadashivanagar, Bengaluru 560080, India}

\preprint{SLAC-PUB-17709}

\begin{abstract}
Nontrivial dark sector physics continues to be an interesting avenue in our quest to the nature of dark matter. In this paper, we study the cosmological signatures of mass-varying dark matter where its mass changes from zero to a nonzero value in the early Universe. We compute the changes in various observables, such as, the linear matter power spectrum and the cosmic microwave background anisotropy power spectrum. We explain the origin of the effects and point out a qualitative similarity between this model and a warm dark matter cosmology with no sudden mass transition. Finally, we do a simple analytical study to estimate the constraint on the parameters of this model from the Lyman-$\alpha$ forest data.
\end{abstract}
 
\maketitle

\section{Introduction}
\label{sec:intro}
Though the presence of dark matter has been  confirmed through its  gravitational effect, the particle nature of DM remains a complete mystery.  Combined with a cosmological constant ($\Lambda$), the simple hypothesis of a cold, collisionless dark matter (CDM) that may or may not interact with ordinary Standard Model (SM) particles is consistent with all cosmological observations to date, on scales ranging from individual galaxies\,\cite{Rubin:1970zza}, to galaxy clusters \cite{Zwicky:1933gu,Clowe:2006eq}, to  cosmological scales as probed by large scale structure\,\cite{Ata:2017dya,DES:2021wwk} and cosmic microwave background (CMB) measurements\,\cite{2016A&A...596A.107P,ACT:2020gnv,SPT-3G:2022hvq}.

During the course of last several decades, myriad  laboratory experiments have been performed to look for any nongravitational interaction of DM. However, none of them so far have yielded any conclusive evidence for its presence. Together they have put stringent limits on the conventional DM theories\,\cite{Schumann:2019eaa}, and have compelled us to theorize novel DM models with nontrivial particle physics phenomena in the dark sector. Such effort have also led us to venture beyond the \emph{vanilla} DM models and design experiments that are better optimized to look for observable signatures such models\,\cite{Battaglieri:2017aum,Berger:2022cab,Cooley:2022ufh,Kahn:2022kae}. 
 On the observation frontier, multiple galactic scale astrophysical anomalies, such as the 
 \emph{diversity problem}\,\cite{Oman:2015xda}, \emph{too big to fail problem}\,\cite{2011MNRAS.415L..40B}, and the \emph{Hubble discrepancy}\,\cite{DiValentino:2021izs,Schoneberg:2021qvd} have raised questions about the simple \lcdm{} model of cosmology, and drawn attention to particle physics models beyond this paradigm.



In this paper we explore one such avenue of nonstandard dark matter physics. We ask the question -- can  the  mass of dark matter particle be dynamical with cosmic time?  In particular, we explore the scenario where DM species was made of massless and hence relativistic particles in the early universe, but after a gradual phase transition at a certain redshift, its constituents acquire mass- eventually forming the CDM population in the Universe. Such a scenario can also be seen as a generalization of two popular cosmological models, namely, the ordinary warm dark matter (WDM) model and the dark radiation model. The former being the limit when the transition happens at a much earlier time, and the latter is when the transition is very late. Myriad of works have studied the nonstandard effects of these cosmological models\,\cite{2013PhRvD..87h3008H,Bernal:2016gxb,DES:2020fxi}.  However, this work broadens the theory space of the cosmological models in a straightforward but interesting way. We also note that a variety of related cosmological models have been studied before such as, late-forming DM\,\cite{Sarkar:2014bca}, ballistic DM\,\cite{Das:2018ons} etc.

The particle physics aspect of mass varying DM (\species) has been explored before in a few works\,\cite{Casas:1991ky,Anderson:1997un,Fardon:2003eh,Chitov:2009ph,Bjaelde:2010vt,Davoudiasl:2019xeb,Mandal:2022yym, Bernardini:2009bn}. 
One example is when the dark matter mass is directly proportional to the vacuum expectation value of a scalar field. The DM mass becomes time-varying when the scalar rolls over a potential\,\cite{Anderson:1997un}. In this model, the scalar vacuum expectation value is inversely related to the DM particle number density, leading to an increase in the DM mass as the Universe expands. A similar  scenario (though in different particle physics context) was introduced for mass varying neutrino models, where a fermionic particle  can have a dynamical mass due to its interaction with a scalar field which may also play the role of the dark energy \cite{Fardon:2003eh, Chitov:2009ph} (also see \cite{Bjaelde:2010vt}). Adopting the same mechanism, recently a mass varying dark matter model was introduced in Ref.\,\cite{Mandal:2022yym}. A fast transition from radiation to matter was discussed in\,\cite{Das:2006ht}. In this case, a phase transition in a very light scalar field sector is responsible for such fast change in equation of state. After the transition, the scalar starts oscillating coherently around a minimum of a quadratic potential and starts behaving like dark matter.
In the above studies, the particle phenomenology was discussed,  but any detailed study on its cosmological implication was missing. Recently, phase space constraint on MVDM were discussed in Ref.\,\cite{Boubekeur:2023fqo}.

In this work, we compute the effects of \species{} in cosmological observables, such as, the linear matter power spectrum and the CMB anisotropy power spectra. We find that the massless phase of \species{} before the transition creates a \emph{suppression} in the linear matter power spectrum that is also reflected in the CMB power spectra.
We explain the origin of the effects and point out a qualitative similarity between this model and a warm dark matter cosmology with no sudden mass transition. As our model deviates from the standard $\Lambda$CDM  scenario at small length scales, we constrain our model using an analytical method with the Lyman-$\alpha$ data from HIRES/MIKE\,\cite{2016A&A...594A..91L,Irsic:2017ixq}. We show that this data already constrains a significant part of the new parameter space. It is important to note that we do not adopt a specific particle physics model but rather focus on a phenomenological model of time variation of the dark matter mass. Our aim is to utilize the known gravitational portal between the dark and the visible sectors of a nonstandard cosmological model without committing to a specific particle physics model. Depending on the specific model however, there might be additional signatures of nonstandard dark sector phenomena which will require dedicated studies.


The outline of the paper is as follows. We describe the background evolution of \species{} and compute the new effects in the background level observable in Sec.\,\ref{sec:bkg}. In Sec.\,\ref{sec:power_spectra}, we solve the Boltzmann equations of the perturbation quantities and compute the changes in the matter and CMB power spectra, followed by statistical comparison of our results with  the SDSS Lyman-$\alpha$ data. 
Finally, We conclude in Sec.\,\ref{sec:conclusion}.

\section{Background evolution}
\label{sec:bkg}
We will assume \species{} to be a fermionic thermal species with a temperature $T$ to keep the model as generic as possible. In this case, the evolution of its background quantities, like energy density, are controlled by its time-varying mass $m(z)$. For the time variation of the mass, we consider a phenomenological model of formation of cold DM of mass $M$ from a massless radiation-like species at a redshift $z_t$. Specifically, we take the following form of $m(z)$ to make the transition between the two epochs smooth,
\begin{equation}\label{eq:m_of_z}
m(z) = \dfrac{M}{2} \left[1-\tanh\left(\dfrac{z-z_t}{\dz}\right)\right]\,.
\end{equation}
Here, $M$ is the final mass of \species, and $\dz$ is the duration of the transition in redshift space. The exact nature and the duration of the transition depend on the underlying particle physics model\,\cite{Bernardini:2009bn,Mandal:2022yym}. In this work, we will only consider fast transition, i.e. $\dz\ll z_t$. The slow transition scenario $\dz\ll z_t$ is qualitatively similar to a WDM model. The actual form of the function $m(z)$ depend on the details of the model too. The form chosen in Eq.(\ref{eq:m_of_z}) is purely phenomenological with minimal number of extra parameters.

\begin{figure}[t]
    \centering
    \includegraphics[width=\columnwidth]{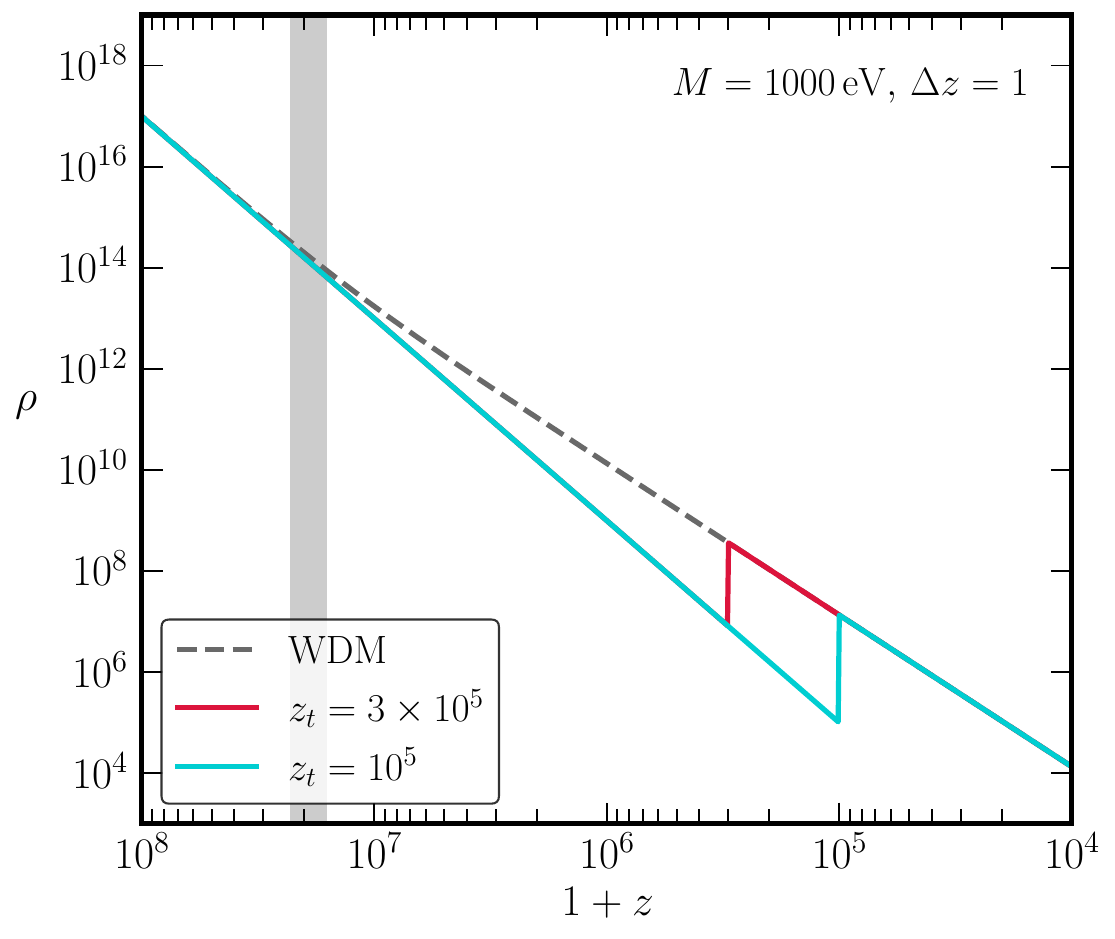}
    \caption{The background density as a function of redshift for mass $M=1\keV$, temperature $T=T_\gamma/10$, and two different transition redshifts $z_t=3\times10^5$ and $10^5$ (red and cyan, respectively). The jumps in the density evolution at $z_t$ are given by $M/T(z_t)=142$ and $426$ in the respective cases. The density evolution of WDM of the same mass is also shown as a dashed line for comparison. The vertical light gray-shaded band shows the time when the WDM particles become nonrelativistic. }
    \label{fig:bkg}
\end{figure}

The phase space distribution $f(q)$ of \species{} is given by a Fermi-Dirac distribution with a temperature $T$. For our
work, we choose $q$ to be the comoving momenta of the particle. We also define $\epsilon=\sqrt{q^2+m(a)^2a^2}$ to be  the energy of the particle; $a=1/(1+z)$ is the scale factor of the Universe. (For choice of momenta and energy see e.g. \cite{Ma:1995ey}). 
As $m(z)$ is zero before the redshift of transition, $z_t$, \species{} behaves as radiation. After $z_t$,  it could behave as radiation or matter depending on its final mass and temperature at that time. It is relativistic if $m(z)/T \ll 3$ during the transition, eventually becoming matter-like, or nonrelativistic if $m(z)/T \gtrsim 3$.
The total energy density of \species{} is given by an energy integral over the phase space distribution:
\begin{equation}
    \rhosp = a^{-4} \int dq~d\Omega~q^2~f(q/T)~\epsilon\,,
\end{equation} 
The energy density at the current epoch is matched to the best-fit energy density of the CDM particle by  Planck. As the particle is non-relativistic at the current epoch $\rhosp \propto T^3 M$. The final energy density of \species{} depends only this combination of mass and temperature and, as discussed later, the evolution of perturbations depends on only $m(z)/T$. This allows us to  express our results in terms of 
the ratio $M/T$. In Fig.\,\ref{fig:bkg}, we show the evolution of the energy density for two different values of $z_t$ that are not excluded by the data. The jumps in $\rhosp(z)$ are given by the quantity $\rhosp^{\rm NR}(z\to z_t^+)/\rhosp^{\rm R}(z\to z_t^-)=M/T_t$ in the respective cases where $T_t$ is the temperature of \species{} at $z=z_t$.\footnote{Note that $\rhosp$ is a continuous function by the construction of $m(z)$ as shown in Eq.(\ref{eq:m_of_z}); only the transition happens very quickly.} These two values of $z_t$ are chosen to show the difference from WDM in the evolution of $\rhosp$. In the rest of the paper, we will fix \species{} temperature to $T=T_\gamma/10$ as a representative value. However, the results are qualitatively same for other lower values of the temperature. The extra period of time \species{} evolves as radiation between the gray vertical band in Fig.\,\ref{fig:bkg} and $z_t$ is the reason behind the new effects on the matter and CMB power spectra as we will see below.

In this work, we will only consider the scenario $M/T_t\gg 1$ which implies that the DM is becomes instantly nonrelativistic when its mass turns nonzero at $z_t$. As discussed before, this choice bridges the gap between WDM and dark radiation models. In passing, we want to compare the present scenario with the ballistic dark matter model considered in Ref.\,\cite{Das:2018ons} which also had a relativistic to nonrelativistic phase transition in the dark sector. However, in that case, the particles were tightly coupled and behaved like a fluid in the relativistic phase that can sustain acoustic waves. Hence it has characteristic features that are distinct from the present noninteracting model.
\begin{figure*}[t]
    \centering
    \includegraphics[width=\columnwidth]{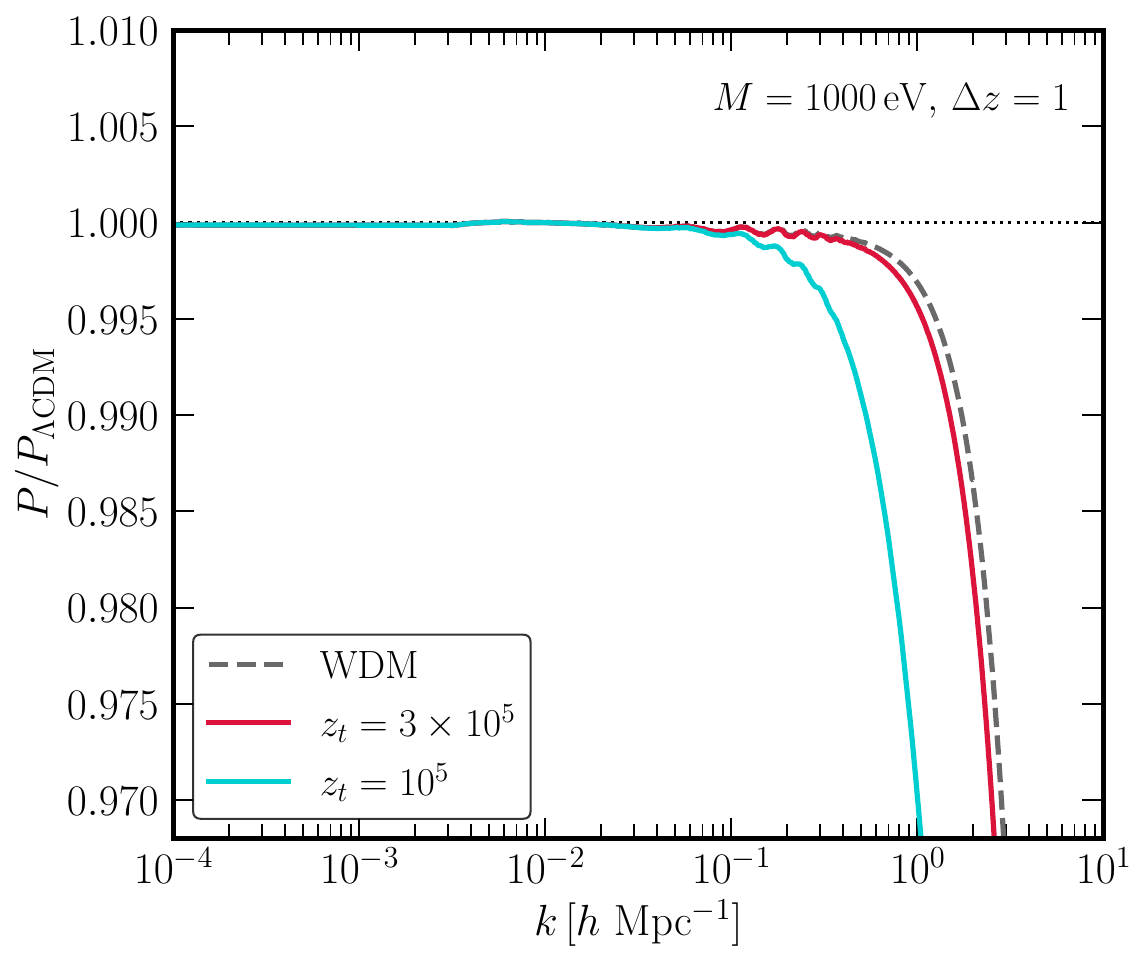}~\includegraphics[width=\columnwidth]{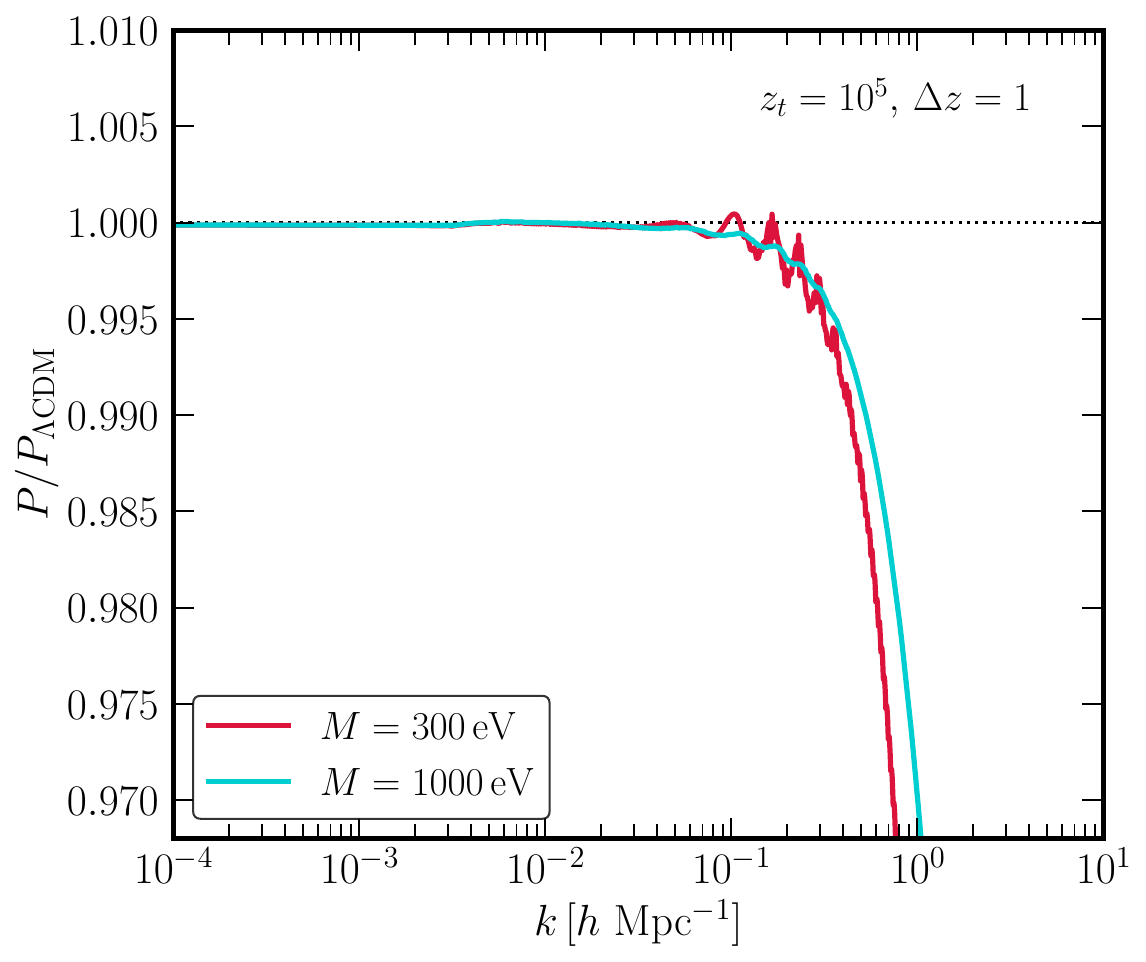}
    \caption{Ratio of the linear matter power spectrum for \species{} to that of \lcdm{}, $P(k)/P_{\Lambda{\rm CDM}}(k)$ for different values of $z_t$ with $M=1000\eV$ (Left), and different values of $M$ with $z_t=10^5$ (Right). See text for the explanation of different features, such as the power suppression at small scales and the wiggles near the cut-off scale. For comparison, we also show the ratio in WDM cosmology with same mass $M$ on the left panel. The temperature is assumed to be $T=T_\gamma/10$ and $\dz=1$.}
    \label{fig:matter_power}
\end{figure*}

\subsection{Extra relativistic degrees of freedom}
In the early Universe before $z_t$, the \species{} acts as radiation and would add to the total relativistic energy density in addition to the photons and neutrinos. This can be quantified as the extra relativistic degrees of freedom $\dneff$ defined as
\begin{equation}\label{eq:neff1}
    \dneff = \dfrac{\rhosp}{\rho^\mathrm{th}_\nu}\,,
\end{equation}
where $\rho^\mathrm{th}_\nu$ is the thermal energy density of a single neutrino species. Because the Universe was radiation dominated before $z\simeq3400$, any extra relativistic energy would have changed the rate of expansion of the Universe. This would have affected the production of light elements during the epoch of big bang nucleosynthesis (BBN) and could also affect the angular power spectra of the fluctuations in the cosmic microwave background (CMB) coming from the epoch of recombination. Very precise observational data from these two era helps us constrain $\dneff$. The fit to the data of the light element abundance in the Universe yields $\neff=2.85\pm0.3$\,\cite{Cyburt:2015mya}. The CMB data from Planck 2018 gives $\neff=2.89^{+0.36}_{-0.38}$ for the TTTEEE+lensing analysis\,\cite{Planck:2018vyg}.

The energy density $\rhosp$ evolves as $\sim(1+z)^3$ after $z_t$, and as $\sim(1+z)^4$ before $z_t$. For a fast transition, we can ignore the duration of the transition. Then the energy density changes by a factor of $M/T_t$. Hence, $\rhosp(z)$ prior to $z_t$ can be written as
\begin{equation}\label{eq:neff2}
    \rhosp(z) \approx \rhosp(z=0)\,\dfrac{T_t}{M}\,\dfrac{(1+z)^4}{1+z_t}\,.
\end{equation}
Here we have normalized the energy density by fixing its today's value $\rhosp(z=0)$ to the observed CDM density $\Omega_\mathrm{c}h^2=0.12$\,\cite{Planck:2018vyg}. For this rough estimate we neglect the current $\Lambda$-dominated epoch because, as we will see later, the $\dneff$ does not yield the strongest bound on this model. Using Eq.(\ref{eq:neff2}) in (\ref{eq:neff1}) then gives
\begin{equation}\label{eq:neff3}
    \dneff \approx \dfrac{\rhosp(z=0)}{\rho^\mathrm{th}_\nu(z=0)}\dfrac{T_t}{M}\,\dfrac{1}{1+z_t}\,,
\end{equation}
during big bang nucleosynthesis (BBN). At present time, neutrino energy density is miniscule relative to DM. In fact, $\rhosp(z=0)/\rho_\nu(z=0)\simeq 10^5$, and, as will be shown shortly, we will mostly consider $z_t\gtrsim10^5$. Therefore, for \species{} that are nonrelativistic at $z_t$ (i.e. $T_t/M\ll1$) will have $\dneff\lesssim 0.01$ and will not contribute significantly to the extra relativistic degrees of freedom in the early Universe during BBN. Moreover, as we are interested in transition epochs much earlier than the recombination and the \species{} density is fixed to the CDM density today, the $\dneff$ by definition vanishes during recombination. Hence, the CMB bound on $\dneff$ does not constrain this model.

\section{Matter and CMB power spectra}
\label{sec:power_spectra}
The radiation phase of \species{} will also affect the evolution of the density fluctuations in the Universe before $z_t$. 
Below, we discuss these effects using the Boltzmann equations in the linear approximation of the perturbations.

The perturbation evolution equations can be obtained from the Boltzmann hierarchy~\cite{Ma:1995ey},
\begin{equation}
\label{eq:boltzmann}
\begin{split}
\dot{\Psi}_0 &= -\dfrac{qk}{\epsilon}\Psi_1 - \dot{\phi}\dfrac{d\ln f_0}{d\ln q}\,,\\
\dot{\Psi}_1 &= \dfrac{qk}{3\epsilon}(\Psi_0-2\Psi_2) - \dfrac{\epsilon k}{3q}\psi\dfrac{d\ln f_0}{d\ln q}\,,\\
\dot{\Psi}_\ell &= \dfrac{qk}{(2\ell+1)\epsilon} \left[\ell\Psi_{\ell-1} - (\ell+1)\Psi_{\ell+1}\right]\,, \quad \ell\ge2\,.
\end{split}
\end{equation}
Here, $\Psi_\ell$ are the $\ell$-th multipole of the perturbations to the phase space distribution function, $\phi$ and $\psi$ are the metric perturbations, $f_0$ is the unperturbed Fermi-Dirac distribution, and $\epsilon=\sqrt{q^2+a^2m(a)^2}$ as mentioned earlier. Macroscopic variables such
as density contrast, bulk velocity, and anisotropic stress can be constructed by integrating Eq.~(\ref{eq:boltzmann}) over comoving momenta. 
The nonvanishing mass prohibits us from performing the phase space integral (over comoving momentum $q$) analytically. Another variable in our study is the unperturbed temperature, $T$,  in the Fermi-Dirac distribution function.   From Eq.~(\ref{eq:boltzmann})  it can be shown that the relevant variables  are $q/T$ and $\epsilon/T$ or, at late times when the particles are non-relativistic, the evolution of the system of equations is determined by $m(z)/T$. We modify the Boltzmann solver code \texttt{CLASS} to add a new species with time-varying mass $m(z)$, and compute the linear matter power spectrum and the CMB anisotropy power spectra~\cite{2011JCAP...07..034B}.

In the left panel of Fig.\,\ref{fig:matter_power}, we show the ratio of the matter power spectra to that in \lcdm{} $P/P_{\Lambda{\rm CDM}}$ for two values of $z_t$ with $\dz=1$ and $M=1000\eV$, and the right panel shows the same ratio for two values of $M$. The main feature in the new power spectrum is the suppression of power at small scale. The \emph{cut-off scale} $k_t$, above which this happens, is the scale that entered the horizon at redshift $z_t$. For a mode $k>k_t$, \species{} was still relativistic when it entered the horizon and was free streaming. This inhibits the growth of structures at those scales resulting in a power suppression. Note that this is qualitatively similar to the analogous feature observed in warm dark matter power spectrum\,\cite{Hu:1997mj,Lesgourgues:2006nd,Viel:2013fqw,TopicalConvenersKNAbazajianJECarlstromATLee:2013bxd}.

This suppression can also be understood from the Boltzmann hierarchy (Eq.~(\ref{eq:boltzmann})), which is  an expansion in $q/\epsilon$. As the particle enters the non-relativistic phase, $q \simeq m a$ and 
$q \ll m a$ at later epochs. This allows us to truncate the hierarchy in Eq.~(\ref{eq:boltzmann}) and obtain the corresponding fluid 
equations (e.g. see Refs.\,\cite{Ma:1995ey,Lesgourgues:2006nd,2010PhRvD..82h9901S}).  The free streaming scale at the time of this transition determines the cut-off scale $k_t$ above which 
the perturbations are wiped out. The  free streaming length is inversely proportional to the thermal velocity  of particles. The thermal velocity is close to the speed of light at the transition and decreases slowly as $1/a$ at later epochs. The comoving free streaming wave number $k_{\rm fs}$ reaches a minimum at the epoch of  transition but continue to be important for much later epochs for perturbations at small scales. In this paper we consider $\Delta z/z_t \ll 1$ or fast transition from a massless to to a massive particle. The models with slow transition could differ significantly from the case we study because of the expected behavior of free streaming scale.  Unlike the WDM model, for which this transition occurs with the onset of the  nonrelativistic era, and hence its perturbations  are  driven solely by the mass of the particle, 
$z_t$ is responsible for the start  of this phase in our case. This is an important distinction between the two models.
\begin{figure*}[t]
    \centering
    \includegraphics[width=\columnwidth]{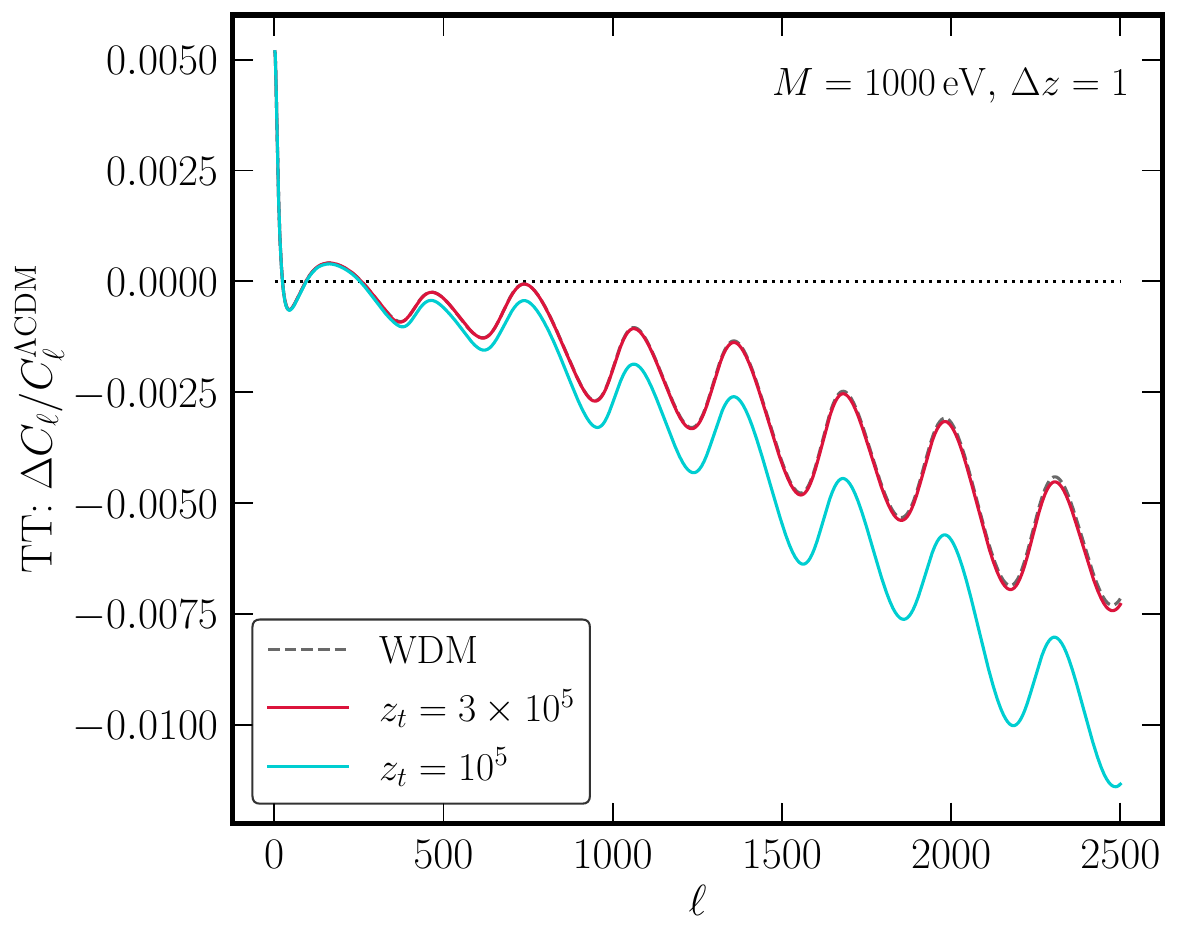}~
    \includegraphics[width=0.98\columnwidth]{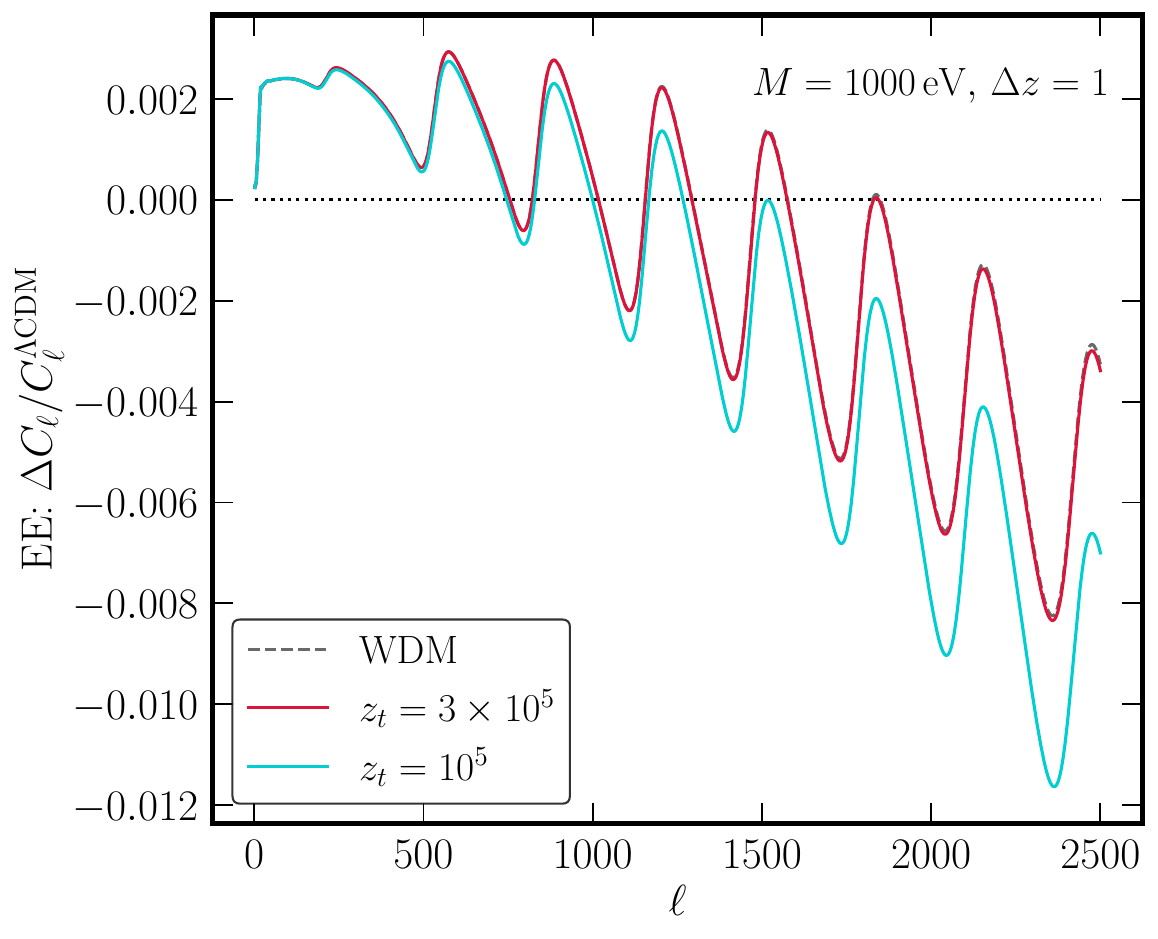}
    \caption{The relative change $\Delta C_\ell/C_\ell^{\Lambda{\rm CDM}}$ of the CMB TT (Left) and EE (Right) angular power spectra for two values of $z_t$ and $M=1000\eV, \dz=1$. See text for the explanation of the oscillatory feature and the power suppression seen here. The corresponding spectrum for the same mass WDM is also shown for comparison. The \species{} spectrum deviates from WDM for lower transition redshift $z_t$.}
    \label{fig:cmb}
\end{figure*}

Later transition means the \species{} was relativistic for longer time, which in turn means they could wash out structures at larger length scales, i.e. smaller $k$ modes. This is evident from the left panel of Fig.\,\ref{fig:matter_power} as the model with $z_t=10^5$ case has a smaller cut-off scale $k_t$ relative to the WDM model with the  same mass. Therefore, it shows that the mass of DM is not the only parameter that controls the cut-off scale, as opposed to the WDM cosmology. From the right panel of Fig.\,\ref{fig:matter_power}, one can see that the cut-off scale also depends on DM mass with smaller $M$ having smaller $k_t$.

Another feature of the ratio of power spectra in Fig.\,\ref{fig:matter_power} is the oscillations around $k\gtrsim0.1\,h\,\mathrm{Mpc^{-1}}$. This is a result of a phase shift in the matter power spectrum relative to \lcdm~\cite{Bashinsky:2003tk,Baumann:2019keh}. Any free streaming relativistic species travels at the speed of light that is greater than the sound speed in the photon-baryon bath before recombination. As a result, they drag the metric perturbations via gravity in a radiation dominated Universe. This in turn creates a phase shift in the acoustic oscillations in the thermal bath manifesting itself in the observable density fluctuations in the Universe today. Such phase shifts have been observed in both CMB anisotropy power spectrum and the baryonic acoustic oscillation (BAO) in the matter power spectrum~\cite{Baumann:2015rya,Follin:2015hya,Baumann:2019keh}. Its presence (or lack of it) has been used to look for other new physics scenarios, including neutrino self-interaction~\cite{Hannestad:2000gt,Friedland:2007vv,Hannestad:2013ana,Archidiacono:2014nda,Kreisch:2019yzn,Das:2020xke,Das:2021guu,Berryman:2022hds,Das:2023npl}. In the present scenario, the \species{} generates this phase shift during its relativistic phase prior to $z_t$ due to its small but nonzero $\dneff$. Because the \lcdm{} matter power spectrum has the BAO oscillations in the range $0.1\lesssim k \lesssim 1\,h\,\mathrm{Mpc^{-1}}$, the phase shift manifests itself in that BAO oscillations\,\cite{Baumann:2019keh}. As expected from Eq.(\ref{eq:neff3}), this oscillatory feature in the matter power spectrum becomes more pronounced (see the right panel of Fig.\,\ref{fig:matter_power}) for lighter mass because it increases $\dneff$ before $z_t$.

The temperature and polarization anisotropy power spectra of the CMB are also modified by the new physics in \species. In Fig.\,\ref{fig:cmb}, we show the relative changes to the TT and EE power spectra. For comparison, we also the spectrum for the WDM model with the same DM mass. The most prominent feature is the power suppression at smaller angular scales which is arises because of the same reason as in the matter power spectrum. This effect is similar to the WDM model. However, the spectrum deviates from the WDM curve for lower $z_t$ as expected because of the longer radiation phase. A phase shift is also present in the CMB spectrum because of the additional $\dneff$ in the \species{} model. It creates the wiggles seen in Fig.\,\ref{fig:cmb}. The wiggles are more pronounced in the EE spectrum as the peaks are sharper compared to the TT spectrum.

\section{Constraints  from Lyman-$\mathbf{\alpha}$ forest data}
\label{sec:constraint}

The  Lyman-$\alpha$ forest  correspond to  the absorption features on the redward side of the rest frame Lyman-$\alpha$ radiation from distant quasars by the intervening neutral hydrogen gas clouds.  The density and amplitude of these features tell us about fluctuations in the neutral hydrogen  density of the  mostly ionized diffuse IGM in the post-reionization era in the redshift range $2.5 \leq z \leq 6$. This in turn tells us about the DM structure formation at those redshifts.  Hydrodynamical simulations have shown that the observed fluctuations correspond to
mildly non-linear density contrast ($\delta < 10$) of the underlying density
field (e.g. \cite{croft1998,croft1999,croft2002,mcdonald,mcquinn2016evolution,2016A&A...594A..91L,Weinberg:2003eg,PASTOR2007411} and references therein). The Lyman-$\alpha$ data allows one to probe the fluctuations of the density field
at scales as small as  the Jeans' scale of the IGM ($k \simeq 5\hbox{--}7 \, \rm Mpc^{-1}$) in the redshift range $2.5 < z < 6$.  The CMB and galaxy data probe much larger scales  $k \lesssim 0.1 \, \rm Mpc^{-1}$\,\footnote{Planck measures the CMB angular power spectrum for $\ell \leq 2500$. The linear scale $k$, corresponding to the angular scale $\ell$, can be computed approximately from $k\eta_0 \simeq \ell$. For the bestfit cosmological model, the conformal time at the current epoch, $\eta_0 \simeq 14000 \, \rm Mpc$.}.  Therefore, the Lyman-$\alpha$ data is particularly suited for our 
study as the  our results deviate  significantly  from the $\Lambda$CDM  (and  WDM) model at small scales (see Fig.\,\ref{fig:matter_power}).  

In this section, we estimate the bounds on the parameters $M$ and $z_t$ using the Lyman-$\alpha$ forest data from HIRES/MIKE spectrographs\,\cite{2016A&A...594A..91L,Irsic:2017ixq}. We use a formalism prescribed in Ref.\,\cite{Schneider:2016uqi,Murgia:2017lwo} to derive the bounds, exploiting the qualitative similarity between the power suppression effects of our model and a WDM model. We essentially compare the area under the 1D matter power spectrum curve with the \lcdm{} model. We compute the 1D matter power spectrum as
\begin{equation}
    \Pd(k) = \frac{1}{2\pi}\int_k^\infty dk'\,k'\,P(k')\,,
\end{equation}
and define the ratio $r(k)$ of the 1D power spectrum w.r.t. that of the \lcdm{} model,
\begin{equation}
    r(k) = \frac{\Pd(k)}{P_{\Lambda{\rm CDM}}(k)}\,.
\end{equation}
Finally, we compare the area under $r(k)$ with a WDM model, that is excluded at 95\% confidence level, to put limit in the $M/T_t-z_t$ plane,
\begin{equation}\label{eq:limit}
    A = \int_{k_{\rm min}}^{k_{\rm max}} dk\,r(k)\,,\qquad \Delta A\equiv1-\frac{A}{A_{\Lambda{\rm CDM}}} < 0.38\,.
\end{equation}
Here, $k_{\rm min}=0.5h\,\mathrm{Mpc}^{-1}$ and $k_{\rm max}=20h\,\mathrm{Mpc}^{-1}$ enclosing the scales that are probed by the Lyman-$\alpha$ data\,\cite{Irsic:2017ixq}. 

\begin{figure}[t]
    \centering
    \includegraphics[width=\columnwidth]{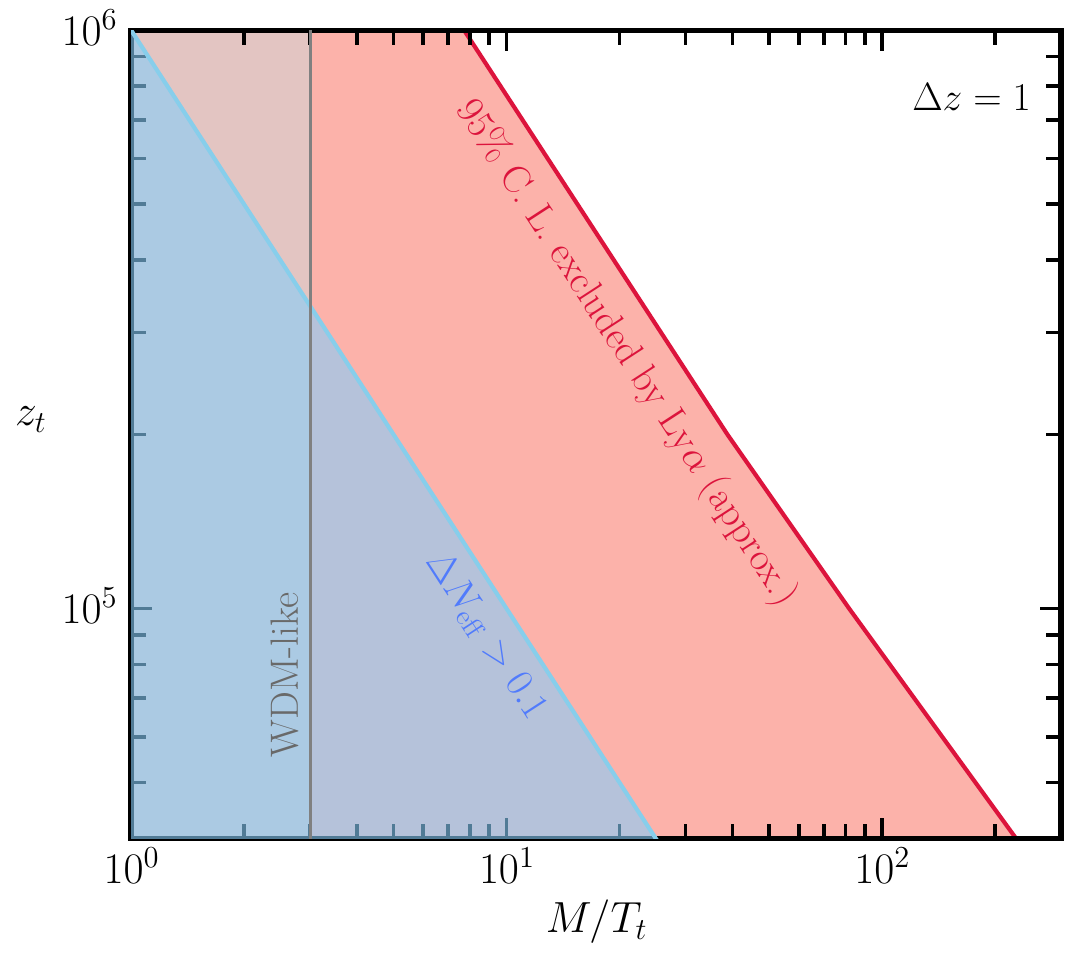}
    \caption{The approximate $2\sigma$ (95\% C.L.) exclusion region in the plane of $M/T_t$ and $z_t$ using the HIRES/MIKE Lyman-$\alpha$ data\,\cite{Irsic:2017ixq}, following the analysis of Refs.\,\cite{Schneider:2016uqi,Murgia:2017lwo}. A fast transition with $\dz=1$ is assumed. The $\dneff$ limit from BBN is also shown as a blue-shaded region. The vertical gray-shaded region on the left is where \species{} is still relativistic when $m(z)$ becomes nonzero and the model is similar to a warm dark matter.}
    \label{fig:limit2}
\end{figure}

In  Fig.\,\ref{fig:limit2} we show  $2\sigma$ (95\% confidence level) exclusion region in the plane of $M/T_t$ and $z_t$ as described above. In the limit of small $M/T$, the \species{} is still relativistic when $m(z)$ becomes nonzero. This scenario is identical to a WDM model (labeled by a gray-shade in Fig.\,\ref{fig:limit2}). The shape of the exclusion region follows  from the discussion in the previous section. Generally for smaller $M/T_t$, the lightness of \species{} and large free streaming length creates more power suppression at small scales than allowed by the data. Whereas, for larger $M/T_t$ the free streaming length decreases, and the parameter space is allowed. We also show the $\dneff$ limit from BBN as a blue-shaded region. The CMB data from Planck 2018 yield a slightly weaker bound. We show only the space above $z_t=4\times10^4$, below which the code \texttt{CLASS} runs into numerical trouble while solving the Boltzmann equations. We believe below this redshift the transition is so delayed that the radiation phase of \species{} before $z_t$ would create large changes in the power spectrum drawing tight constraints from both CMB and Lyman-$\alpha$ data. We plan to investigate this part of the parameter space further in a future work. 

The limit $\Delta A<0.38$ used in Eq.(\ref{eq:limit}) is approximate, and based on an earlier analysis corresponding to a WDM mass $m_\mathrm{WDM}=3.5\keV$. More recent studies, using both Lyman-$\alpha$ and other methods, have yielded different values for the lower limit\,\cite{Garzilli:2019qki,Palanque-Delabrouille:2019iyz,DES:2020fxi,Enzi:2020ieg,Hooper:2022byl,Villasenor:2022aiy} (including one with a weak preference for a nonzero WDM mass\,\cite{Villasenor:2022aiy}). However, we note that the differences between these results are not large enough to drastically affect our results.  Finally, we note that a more comprehensive comparison of our model with the Lyman-$\alpha$ data will require a statistical analysis which we leave for a future work.

\section{Conclusion} 
\label{sec:conclusion}
We have studied the cosmological signatures of a mass-varying dark matter (MVDM) model where the mass $m(z)$ of the DM particle changes from zero to $M$ at a redshift $z_t$. We consider scenarios in which the transition is fast, i.e. $\Delta z/z \ll 1$ and the \species{} is nonrelativistic when $m(z)$ becomes nonzero. We have computed the linear matter power spectrum and the CMB angular power spectra in this model.

The main new effects on the matter and CMB power spectra are power suppression at small scales and a phase shift in the BAO and CMB peaks.
These effects are qualitatively similar to that of a WDM model. In both cases, the free streaming of DM with large thermal velocity impedes structure formation at small scales resulting in  matter power suppression. However, relative to a WDM of same mass, \species{} yields greater suppression and impact larger scales (or a smaller cut-off scale $k_t$). This is because the particle is massless before the transition redshift $z_t$. Therefore, unlike the WDM model,  both the mass $M$ and transition redshift $z_t$ determine the relevant scales at which the power is suppressed and the amount of suppression. 

Similarly, the CMB temperature and polarization anisotropy power spectra receive changes in this model that are broadly similar to the WDM case but distinct from it. In addition to the suppression of power, there is a phase shift in the oscillations of the CMB spectra. This could constitute a tell-tale signature of any new radiation and/or secret interaction in the early Universe\,\cite{Bashinsky:2003tk,Baumann:2015rya}. Near future CMB experiments, that plan to measure the polarization anisotropies more precisely, can attempt  to detect such phase shift in the power spectra.

As the expected departure of our model from the standard  cases is more prominent at small scales, we constrain this model using the Lyman-$\alpha$ data. In the current work, we do not attempt to pinpoint the unique features of our model. 
The degeneracy between $M$ and $z_t$  might make this task difficult. For example, a small scale power suppression might be explainable  either  by WDM or \species{} of larger mass but later $z_t$. Further work is needed to find any qualitative differences between the two models and quantify them. In a future work, we plan to statistically analyse   this model in more detail, and quantify its differences
from a WDM model.

\acknowledgments
We thank Ethan Nadler for helpful comments on the manuscript. The work of A.D. was supported by the U.S. Department of Energy under contract number DE-AC02-76SF00515 and Grant Korea NRF-2019R1C1C1010050. A.D. gratefully acknowledges the hospitality at APCTP during the completion of this work. SD acknowledges SERB DST Government of India
grant CRG/2019/006147 for supporting the project.

\bibliographystyle{apsrev4-2}
\bibliography{main}


\end{document}